\documentclass[usenatbib]{mnras}

\usepackage[final]{microtype}
\usepackage[T1]{fontenc}
\usepackage{pifont,stmaryrd} 
\usepackage{ae,aecompl}
\usepackage{graphicx}
\usepackage{amsmath}
\usepackage{diagbox}
\usepackage{amssymb}	
\usepackage{subfig}
\usepackage{multirow}
\usepackage{tabularx}
\usepackage[mathscr]{euscript}
\usepackage{pifont}
\usepackage{array}
\usepackage[usenames,dvipsnames,table]{xcolor}
\usepackage{braket}
\usepackage{enumerate}
\usepackage{amsfonts}
\usepackage{scalefnt}
\usepackage[normalem]{ulem}
\usepackage{tikz}   
\usepackage{float}
\restylefloat{table}
\newcolumntype{C}{>{$}c<{$}}
\usepackage{booktabs} 
\newcolumntype{P}[1]{>{\centering\arraybackslash}p{#1}}
\def\simless{\mathbin{\lower 3pt\hbox
{$\rlap{\raise 5pt\hbox{$\char'074$}}\mathchar"7218$}}}   
\def\simmore{\mathbin{\lower 3pt\hbox
{$\rlap{\raise 5pt\hbox{$\char'076$}}\mathchar"7218$}}}   
\newcommand{\be}{\begin{equation}}
\newcommand{\ee}{\end{equation}}
\def\fedd{{\rm f}_{\rm Edd}}
\newcommand{\tgr}{{\psi}_{\textsc{GR}}}
\newcommand{\tenv}{\psi_{\textsc{GR+Env}}}
\newcommand{\tecc}{\psi_{\textsc{GR+Ecc}}}
\newcommand{\ttgr}{\psi_{\textsc{GR+TGR}}}
\newcommand{\tenvtgr}{\psi_{\textsc{GR+Env+TGR}}}
\newcommand{\tecctgr}{\psi_{\textsc{GR+Ecc+TGR}}}
\newcommand{\hth}{\hat{\theta}}
\newcommand{\MSun}{{\rm M}_{\sun}}

\title[Systematics in TGR using LISA MBHBs]{Systematics in tests of general relativity using LISA massive black hole binaries}

\author[M. Garg et al.]{Mudit Garg,$^{1}$\thanks{E-mail: mudit.garg@uzh.ch} Laura Sberna,$^{2}$ Lorenzo Speri,$^{3}$ Francisco Duque,$^{4}$ and Jonathan Gair$^{4}$\\
$^{1}$Department of Astrophysics, University of Zurich, Winterthurerstrasse 190, CH-8057 Z\"urich, Switzerland\\
$^{2}$ School of Mathematical Sciences, University of Nottingham, University Park, Nottingham NG7 2RD, United Kingdom \\
$^{3}$ European Space Agency (ESA), European Space Research and Technology Centre (ESTEC), Keplerlaan 1, 2201 AZ Noordwijk, the Netherlands\\
$^{4}$Max Planck Institute for Gravitational Physics (Albert Einstein Institute) Am Mühlenberg 1, D-14476 Potsdam, Germany\\
}

\date{Received / Accepted}
\pubyear{2024}
\begin{document}
\label{firstpage}
\pagerange{\pageref{firstpage}--\pageref{lastpage}}
\maketitle

\begin{abstract}
Our current understanding is that an environment -- mainly consisting of gas or stars -- is required to bring massive black hole binaries (MBHBs) with total redshifted mass $M_z\sim[10^{4},10^7]~\MSun$ to the LISA band from parsec separation. Even in the gravitational wave (GW) dominated final inspiral, realistic environments can non-negligibly speed up or slow down the binary evolution, or leave residual, measurable eccentricity in the LISA band. Despite this fact, most of the literature does not consider environmental effects or orbital eccentricity in modelling GWs from near-equal mass MBHBs. 
Considering either a circular MBHB embedded in a circumbinary disc or a vacuum eccentric binary, we explore if ignoring either secular gas effects (migration and accretion) or eccentric corrections to the GW waveform can mimic a failure of General Relativity (GR). We use inspiral-only aligned-spin 3.5 post-Newtonian waveforms, a complete LISA response model, and Bayesian inference to perform a parameterized test of GR. For a four-year LISA observation of an MBHB with $M_z=10^{5}~\MSun$, primary-to-secondary mass ratio $q=8$, and component BHs' dimensionless spins $\chi_{1,2}=0.9$ at redshift $z=1$, even a moderate gas-disc imprint (Eddington ratio $\fedd\sim0.1$) or low initial eccentricity ($e_0\sim10^{-2.5}$) causes a false violation of GR in several PN orders. However, correctly modelling either effect can mitigate systematics while avoiding significant biases in vacuum circular systems. The adoption of LISA makes it urgent to consider gas imprints and eccentricity in waveform models to ensure accurate inference for MBHBs.
\end{abstract}

\begin{keywords}
accretion, accretion discs -- black hole physics -- gravitational waves -- methods: data analysis -- methods: statistical.
\end{keywords}

\section{Introduction}

The adoption of the Laser Interferometer Space Antenna (LISA; \citealt{AmaroSeoane2017,Colpi2024}) by the European Space Agency opened up an exciting opportunity to observe gravitational waves (GWs) in the milliHz band, together with in-development projects TianQin \citep{Wang2019,Li2024} and Taiji \citep{Gong2021}. One of the primary sources for LISA is massive black hole binaries (MBHBs) with total redshifted mass $M_z\sim[10^4,10^7]~\MSun$ and primary-to-secondary mass ratios $q\lesssim10$. LISA will observe MBHBs up to redshift $z\sim20$, with up to a few years of inspiral before merging, and high signal-to-noise ratios (SNRs; \citealt{AmaroSeoane2017}), as large as $\mathcal{O}(1000)$. These properties will allow to not only constrain source parameters but help to understand the evolution of massive BHs and may also put bounds on their astrophysical environment \citep{Barausse2014,Garg2022,Garg2024b,Garg2024c,Tiede2024b,Zwick2024}, and further constrain deviations from General Relativity (GR) using GWs \citep{TGR_GWTC1,TGR_GWTC2,TGR_GWTC3}.

MBHBs mainly form as a by-product of galaxy mergers \citep{Begelman1980}. Following a merger, dynamical friction can bring a MBH pair to the center of a newly formed galaxy to form a bound MBHB at a parsec scale. The binary then typically needs to be further driven by gas or stars to reach milli parsec separation, where GW emission can take over and finally bring the MBHB to the LISA band \citep[see, e.g.][]{Sesana2011,AmaroSeoane2022} in a Hubble time. Radiatively efficient gas can settle down as a circumbinary disc (CBD) around the MBHB \citep{DOrazio2016} and even in the GW-dominated regime, it can remain coupled to the binary until near-merger \citep{Dittmann2023, Gutierrez2024}. CBDs can slow-down or speed-up the binary inspiral \citep{Barausse2014,Garg2022,Dittmann2023,Tiede2024b,Garg2024b,Zwick2024}, sustain measurable residual orbital eccentricity \citep{Roedig2011,Zrake2021,DOrazio2021,Siwek2023,Garg2024a}, align the component BHs' spins to the binary angular momentum and alter their magnitudes \citep{Bardeen1970,Thorne1974,Bardeen1975}, and also produce electromagnetic (EM) counterparts \citep{Haiman2017,dAscoli2018,Mangiagli2022,Cocchiararo2024}. Even in the absence of a CBD, residual eccentricity in the LISA band can be due to stellar hardening \citep{Quinlan1996,Khan2011,Gualandris2022}, interactions with a third MBH \citep{Blaes2002, Naoz_2016, Bonetti2019}, or a binary inspiral taking longer than the AGN lifetime of $\sim100$ Myr \citep{Hopkins2009}, such that gas dissipates much before merger. 

Despite these expectations of non-vacuum or non-circular perturbations to the GW waveform, current literature mostly ignores them for MBHBs. However, if we have either significant gas presence, accreting with Eddington ratio $\fedd\gtrsim0.1$, or non-negligible residual eccentricity ($e\gtrsim10^{-2.5})$ during the binary inspiral in the LISA band, then not modeling them would not only remove formation channel signatures but could bias intrinsic binary parameters \citep{Garg2024b}, leading to an inaccurate recovered MBH mass function as one major consequence. Another repercussion of ignoring environmental effects or eccentricity could be systematics in tests of GR using GWs from MBHBs. In observations by the ground-based LIGO-Virgo-KAGRA (LVK) collaboration, there have been reports \citep{Romero-Shaw2020,Romero-Shaw2022,Gupte2024} of stellar-mass BHB signals with non-zero eccentricity, although there is not yet general consensus within the community \citep{LVKEcc2024}. Furthermore, there have been works that explored how ignoring eccentricity and other environmental effects can cause systematics in testing GR \citep{Saini2022,Narayan2023,CanevaSantoro2024,Gupta2024}. This was also shown for extreme mass ratio inspirals \citep{Speri:2022upm, Kejriwal:2023djc} and the same could hold for LISA MBHBs.

In this work, we explore how ignoring either gas-induced secular dephasing of the GW waveform or astrophysically-motivated initial orbital eccentricity, while modeling the GW inspiral via Post-Newtonian (PN) expansion, can lead to false deviations from GR and how we can mitigate these systematics. To closely mimic realistic LISA data analysis, we employ high-order inspiral-only aligned-spin-corrected PN waveforms, a complete LISA response model, and Bayesian inference. 

The paper is structured as follows. In Section~\ref{S:waveform}, we present the relevant GW waveform templates and parameter space. Section~\ref{S:analysis} explains our analysis methodology to quantify deviations from GR. In Section~\ref{S:results}, we show our results for either gas-rich or vacuum eccentric MBHBs. We discuss our findings in Section~\ref{S:discussion} and conclude in Section~\ref{S:conclusion}.

\section{Waveform templates, parameters of interest, and LISA response}\label{S:waveform}

We consider a MBHB with total redshifted binary mass $M_z$ at redshift $z$, primary-to-secondary mass ratio $q\lesssim10$, and aligned component BHs' dimensionless spins $\chi_{1,2}$ in an orbit with detector-frame semi-major axis $a$. The GW phase of the dominant harmonic of this system in the stationary phase approximation\footnote{We always model our amplitude as Newtonian, without any corrections. This is for simplicity and due to the expectation that small non-vacuum,  non-circular corrections affect phase more sensitively than amplitude \citep{Moore2016}.} (SPA; \citealt{Cutler1994}) for a circular orbit in vacuum is our GR template ($\psi_{\rm GR}$). It is expressed as
\begin{equation}\label{eq:phaseGR}
    \psi_{\rm GR}=2\pi f t_c-\phi_c+\frac{3}{128\eta v^{5}}\sum_{n}\left(\psi_{n}+\psi_{n}^{(l)}\log v\right)v^{n},
\end{equation}
where $t_c$ and $\phi_c$ are the time and phase at coalescence, respectively, $f=(1/\pi)\sqrt{GM_z/a^3}$ is the observed GW frequency of the $(2,2)$ (quadrupole) mode,\footnote{This should also be the case for later considered weak effective gas amplitudes and small eccentricities.} $v\equiv(GM_z\pi f/c^3)^{\frac13}$ and $\eta\equiv q/(1+q)^2$ are the characteristic velocity and the symmetric mass ratio of the binary, respectively. Moreover, the index $n$ represents the $n/2$-th relative PN order, and $\psi_n$ are the GR coefficients for $n\in\{0,2,3,4,5,6,7\}$ and $\psi^{(l)}_n$ are the logarithmic GR coefficient for $n\in\{5,6\}$ in vacuum, with aligned spin corrections up to $3.5$ PN order. Both $\psi_n$ and $\psi^{(l)}_n$ are explicit functions of $\eta$ and $\chi_{1,2}$ \citep{Buonanno2009,Arun2009,Mishra2016}.

Radiatively efficient gas around a MBHB can settle into a circumbinary disc (CBD; see, e.g., \citealt{Escala2004,DOrazio2016} and the review \citealt{Lai:2022ylu}) and can induce a perturbative correction to $\tgr$ due to binary migration and accretion of gas onto the MBHB. Assuming a circular inspiral, using the SPA, and keeping only the leading order in the environmental effect, the phase contribution due to the interaction with CBD takes the form \citep{Yunes2011,Garg2024b}
\begin{equation}
    \psi_{\rm GR+Env}=\psi_{\rm GR}+\frac{3}{128\eta v^{5}}A_{\rm g}v^{-2n_{\rm g}}\left(\frac{\eta}{0.1}\right)^{-2}\left(\frac{M_z}{10^5\MSun}\right),
\end{equation}
where $n_g$, typically $>0$, sets the power-law dependence of the environmental effect on the semi-major axis, and $A_g$ is the effective amplitude of the environmental effect,
\begin{equation}
    A_g= 10^{-14}\left(\frac{\xi}{100}\right)\left(\frac{{\rm f}_{\rm Edd}}{0.1}\right)\left(\frac{0.1}{\epsilon}\right), 
\end{equation}
which depends only upon disc parameters: the Eddington ratio ${\rm f}_{\rm Edd}$, radiative efficiency $\epsilon$, and the strength of the torque $\xi$ exerted by the gas disk, which depends sensitively on the binary mass ratio, other disc parameters, and the disk thermodynamics \citep{Dittmann2022,Garg2022,Duffell2024}. Most current hydrodynamical simulations  -- performed in a sub-parsec regime assuming a fixed Newtonian orbit -- suggest that $|\xi|\lesssim10$.  Recently, \cite{Garg2024e} reported $\xi\sim-36$ in the LISA band, by coupling the PN live-orbit binary inspiral under GWs to the gas-driven evolution in hydrodynamics simulations. However, we normalize $\xi$ by $100$ as all numerical studies performed have some missing physics, such as magnetic fields or radiative transfer. Therefore, torque can become stronger or even weaker when properly estimated in a complete binary-disc interaction study. Note that a positive (negative) $A_g$ increases (decreases) the GW phase by slowing-down (speeding-up) the binary inspiral. We only consider secular gas effects from CBDs -- migration and accretion -- which typically have $n_g=4$ \citep{Caputo2020,Garg2024b}.

Similarly, we include low-eccentricity SPA corrections to our circular vacuum template to build our vacuum eccentric template
\begin{equation}
    \psi_{\rm GR+Ecc}=\psi_{\rm GR}-\frac{3}{128\eta v^{5}}\frac{2355}{1462}\left(\frac{v_0}{v}\right)^{\frac{19}{3}}e_0^2\sum^{6}_{n=0}\psi^{\rm ecc}_{n}v^{n},
\end{equation}
where $e_0$ is the initial eccentricity (measured at the start frequency, related to $v_0$, of the LISA observation for the given event) and $\psi^{\rm ecc}_{n}$ are explicit functions of $\eta$ \citep{Moore2016}. Since, spin corrections to the GR phase in Eq.~\eqref{eq:phaseGR} enter at $1.5$PN order (spin-orbit coupling) and beyond, and we consider small eccentricity in this work, we reasonably expect any possible spin-eccentricity cross-terms to be unimportant. 

We are interested in determining whether ignoring either gas corrections or eccentricity can mimic a statistically significant violation of GR in the GW signal of a binary inspiral. To this aim, we introduce a set of test-of-GR (TGR) parameters \citep{Li2012} $\delta\psi_k$ and $\delta\psi^{(l)}_k$ (for logarithmic terms) in the phase of our previously introduced models:
\begin{align}\label{eq:phaseTGR}
    \psi_{\rm Model + TGR}=&\psi_{\rm Model}+\\
    &\frac{3}{128\eta v^{5}}\sum_{k}\left(\psi_{k}\delta\psi_k+\psi_{k}^{(l)}\delta\psi^{(l)}_k\log v\right)v^{k},\nonumber
\end{align}
where $k\in\{-2,0,1,2,3,4,6,7\}$ for $\delta\psi_k$ and $k\in\{5,6\}$ for $\delta\psi^{(l)}_k$. Since GR does not have $-1$PN and $0.5$PN terms, the TGR parameters at these orders need to be interpreted as absolute errors, and we subsequently replace $\psi_{-2}\delta\psi_{-2}\to\delta\psi_{-2}$ and  $\psi_{1}\delta\psi_{1}\to\delta\psi_{1}$.

In Table~\ref{table:parameters}, we list our parameters of interest and their values in a reference system. We chose the reference system primarily to have both high SNR as well as numerous GW cycles in LISA. We consider highly spinning ($\chi =0.9$) MBHs because we expect prograde CBDs, resulting in prograde minidiscs \citep{DOrazio2021}, to spin up the MBHs to almost unity dimensionless spin by the time the MBHB reaches the LISA band \citep{Garg2024c}. This is further supported by the fact that most observed AGNs have highly spinning central MBHs \citep{Reynolds2021}. Also, having different fiducial spins does not affect our results, as discussed in \S~\ref{S:EnvComp}. We further select $q=8$ as our fiducial value, which leads to thousands of GW cycles, because, even if the MBHB started at a parsec scale with a highly asymmetric mass ratio, expected preferential accretion onto the secondary MBH is likely to bring it to a near-equal mass ratio a few years before merger \citep{Duffell2020}. However, it may not be precisely one as it will depend on the inspiral timescale, AGN lifetime \citep{Hopkins2009}, and steadiness of accretion onto component MBHs. We also discuss the case of almost unity mass ratio in Figs.~\ref{fig:MCMCEnvComp} and \ref{fig:MCMCEccComp}. 

We dub $\{M_z,q,\chi_{1,2},t_c,A_g,n_g,e_0,\delta\psi_k,\delta\psi^{(l)}_k\}$ as intrinsic parameters and the rest of the parameters as extrinsic. The initial eccentricity $e_0=10^{-2.5}$ is the expected one if the binary is driven by prograde accretion from the parsec scale  at $\fedd=1$ \citep{Zrake2021,Garg2024c}. Higher initial eccentricities could be induced by retrograde CBDs \citep{Garg2024c} or interaction with a third MBH \citep{Bonetti2019}. Moreover, we term $\{M_z,q,\chi_{1,2},t_c\}$ as intrinsic-merger and $\{A_g,n_g,e_0\}$ as intrinsic-inspiral parameters due to their relative importance in the inspiral.

\begin{table}
\centering
    \begin{tabular}{|C|p{0.45\linewidth}|C|}
        \hline
        \pmb{\theta}&{\bf Definition} & {\rm \bf Fiducial}\\
        \hline
        \hline
        M_z &Total redshifted mass & 10^5~\MSun\\
        \hline
        q~(\eta) & Mass ratio (symmetric mass ratio)& 8.0~(0.1)\\
        \hline
        \chi_{1,2}& Dimensionless spin parameters for component BHs& 0.9\\
        \hline
         t_c &Time of coalescence& 4~{\rm years}\\
        \hline
        A_{\rm g}&Environmental amplitude  & 10^{-14}\\
        \hline
        n_{\rm g} & Environmental semi-major axis power-law relative to GWs & 4\\
        \hline
        e_0 & Initial orbital eccentricity & 10^{-2.5}\\
        \hline
        \delta\psi_k,\delta\psi^{(l)}_k& Test of GR parameters & 0\\
        \hline
        D_{\rm L}~(z)&Luminosity distance (redshift) & 6791.3~{\rm Mpc}~(1)\\
        \hline
        \imath & Inclination & 0.5~{\rm rad} \\ 
        \hline
        \phi_c&Phase at coalescence & 0.5~{\rm rad}\\ 
        \hline
        \phi& Observer's azimuthal phase & 0.5~{\rm rad}\\ 
        \hline
        \lambda  & Ecliptic latitude & 0.5~{\rm rad} \\ 
        \hline
         \beta&Ecliptic longitude & 0.5~{\rm rad} \\ 
        \hline
         \psi&Initial polarization angle & 0.5~{\rm rad} \\ 
        \hline
    \end{tabular}
\caption{Parameters of our reference system in the LISA frame. This system has SNR of $\sim 378$ and $\sim 19000$ GWs cycles.}
\label{table:parameters}
\end{table}

We add all considered waveform templates to the package \textsc{lisabeta} \citep{Marsat2021} to produce a complete LISA response, including the motion of LISA and time delay interferometry (TDI; \citealt{Tinto2021}). We generate waveforms for MBHBs until their respective innermost stable circular orbit.

In the next section, we describe how we use Bayesian inference to quantify deviations from GR.

\section{Analysis method}\label{S:analysis}

We perform a Bayesian analysis on the following parameters, depending on the waveform template (superscript):
\begin{align}\label{eq:recov}
    \theta^{\textsc{GR}}_{\rm rec}&=\{M_z,q,\chi_1,\chi_2,t_c\},\\
    \theta^{\textsc{GR+Env}}_{\rm rec}&=\theta^{\textsc{GR}}_{\rm rec}\cup\{A_g\},\nonumber\\
    \theta^{\textsc{GR+Ecc}}_{\rm rec}&=\theta^{\textsc{GR}}_{\rm rec}\cup\{e_0\},\nonumber\\
    \theta^{\textsc{GR+TGR}}_{\rm rec}&=\theta^{\textsc{GR}}_{\rm rec}\cup\{\delta\psi_k\}\nonumber,\\
    \theta^{\textsc{GR+Env+TGR}}_{\rm rec}&=\theta^{\textsc{GR}}_{\rm rec}\cup\{A_g,\delta\psi_k\}\nonumber,\\
    \theta^{\textsc{GR+Ecc+TGR}}_{\rm rec}&=\theta^{\textsc{GR}}_{\rm rec}\cup\{e_0,\delta\psi_k\}\nonumber,
\end{align}
with $\delta\psi^{(l)}_k$ replacing $\delta\psi_k$ for logarithmic terms. Because we limit our analysis to the inspiral portion of the signal, we choose to fix the phase at coalescence $\phi_c$. Inferring $\phi_c$ with an inspiral-only template would lead to inaccurate posteriors for some TGR parameters, due to a partial degeneracy between higher order TGR parameters with index $k\gtrsim4$ and $\phi_c$, as discussed in Appendix~\ref{App:phic}. The latter leads to non-Gaussian and even almost flat, uninformative posteriors for those TGR coefficients. 
The inclusion of either higher modes due to the unequal mass ratio of the system or the merger-ringdown part of the signal typically breaks this degeneracy. We find that our results are consistent with the inspiral-merger-ringdown analysis of \citet{Chamberlain2017} (for their choice of MBHB: $M_z=10^{7.5}~\MSun,~q=1.2,~\chi_1=0.9,~\chi_2=0.7,~t_c=3~{\rm years}$, and $z=5$). 
We also always fix the slope of the environmental effect $n_g$ -- equivalent to assuming good knowledge of the dependence of the environmental effect on the orbital separation -- as well as the extrinsic parameters to their fiducial values.\footnote{\citealt{Garg2024b} found that inferring $n_g$ leads to non-Gaussian and biased posteriors for the intrinsic-inspiral parameters, due to the low SNR in the early inspiral, coupled with a degeneracy between $A_g$ and $n_g$. Inference on the extrinsic parameters leads to slightly wider posteriors for all parameters.}

For Bayesian inference, we use the same setup as \citet{Garg2024a,Garg2024b}: we employ the $\textsc{PTMCMC}$ sampler,\footnote{https://github.com/JohnGBaker/ptmcmc} uniform priors for all parameters, zero noise realization, and Fisher initialization.\footnote{See Appendix~\ref{App:Bayesian} for priors on different parameters and further explanation on our Bayesian setup.}

To determine if we have a statistically significant deviation from GR due to unmodeled non-vacuum or non-circular effects, we compute the absolute biases on the recovered parameters, which we define for the $i$th parameter as
\begin{equation}\label{eq:Bias}
    \Delta \theta^i=|\bar{\theta}^i_{\rm rec}-{\theta}^i_{\rm inj}|,
\end{equation}
where ${\theta}^i_{\rm inj}$ is the injected value and $\bar{\theta}^i_{\rm rec}$ is the median of the respective posterior, corresponding to the maximum likelihood value in the limit of a Gaussian posterior. We then compute the fractional bias, obtained by dividing the absolute bias by the parameter's statistical error $\sigma_\theta$ (standard deviation of the posterior, assuming it is Gaussian):
\begin{equation}\label{eq:fracBias}
    \hth=\frac{\Delta \theta}{\sigma_\theta}.
\end{equation}
We consider the bias to be significant when $\hth>2$, i.e, beyond two-sigmas.

We also compare TGR models with GR models by evaluating Bayes factors. This is the ratio between the evidence $Z$ of a TGR template (X+TGR) and a respective non-TGR template (X) for a given injected signal,
\begin{align}\label{eq:BayesF}
    \mathcal{B}=\frac{Z_{\rm X+TGR}}{Z_{\rm X}},
\end{align}
where X can be GR, GR+Env, or GR+Ecc. For $|\ln\mathcal{B}|>5$, we have decisive evidence in favor of the TGR (GR) template for positive (negative) $\ln\mathcal{B}$ \citep{Taylor2021}. Otherwise, we have inconclusive evidence in favor of either template. We estimate the evidence using the stepping stone algorithm \citep{Xie2010,Karnesis2023}.

In the next section, we analyze several scenarios to assess whether ignoring either gas-induced corrections or eccentricity can lead to (apparent) statistically significant violation of GR. 

\section{Results}\label{S:results}
We consider three types of signals (injections): vacuum circular ($\tgr$), including gas-induced perturbation ($\tenv$), or having orbital eccentricity ($\tecc$). We then recover them with TGR templates. This procedure quantifies whether ignoring non-vacuum or non-circular effects during Bayesian inference leads to false deviations from GR at a given PN order, i.e., $\delta\hat\psi_k>2$. We further show Bayes factors in favor of the given TGR template over the corresponding non-TGR template when fitting an injected signal as given in Eq.~\eqref{eq:BayesF}. We always recover almost Gaussian posteriors, and all Bayes factors have $\lesssim0.1$ errors. For each case, we first study our fiducial system and then explore the wider parameter space by changing one analysis aspect at a time. 

\subsection{Gas-rich circular systems}
\subsubsection{Fiducial MBHB}
In this section, we inject either a vacuum circular template ($\tgr$) or an environmental circular template ($\tenv$) with two different effective gas amplitudes $A_g\in\{10^{-14},10^{-13}\}$. Then we recover them with a vacuum circular TGR template ($\ttgr$) and an environmental circular TGR template ($\tenvtgr$). Both values of $A_g$ are measurable by LISA for our fiducial MBHB \citep{Garg2024b}, i.e., they can be recovered with relative error $\sigma_{A_g}/A_g<50$ per cent, which signifies more than two-sigma confidence in a non-zero value.

\begin{figure}
    \centering
    \includegraphics[width=0.5\textwidth]{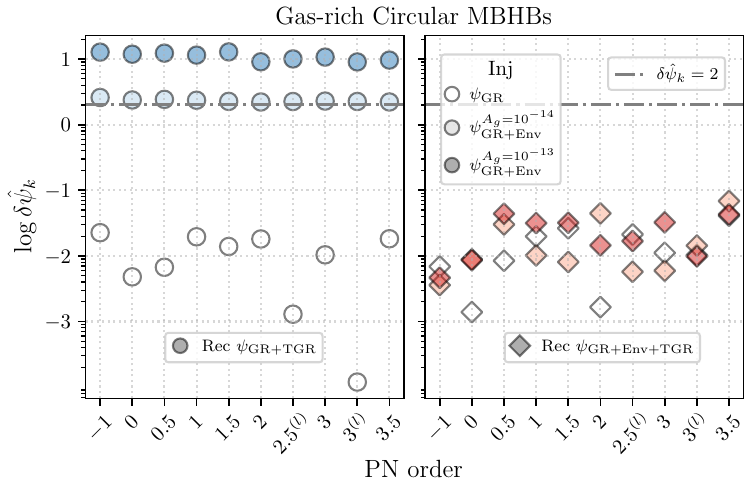}
    \caption{
    Fractional bias induced on TGR parameters by gas-rich signals. We consider three injections: vacuum circular GR ($\tgr$; no fill), and circular GR with a small environmental effect ($A_g=10^{-14}$; light fill) and a strong environmental effect ($A_g=10^{-13}$; dark fill). The biases indicate significant deviations ($\delta\hat\psi_k>2$, above dot-dashed line) from GR at all PN orders for either environmental amplitude when the template neglects the effect ($\ttgr$, circles, left panel). Including the environmental perturbation in the TGR template ($\tenvtgr$, diamond, right panel) mitigates this systematic bias. 
    All signals are from MBHBs with $M_z=10^5~\MSun,~q=8,~\chi_{1,2}=0.9$, and $t_c=4$ years.
    }
    \label{fig:MCMCEnvBias}
\end{figure}

In Fig.~\ref{fig:MCMCEnvBias}, we show the fractional biases on the TGR parameters ($\delta\hat\psi_k$). As expected,  we find that, when analysing a vacuum GR signal with a GR+TGR template, we recover TGR parameters consistent with their injected zero value (left panel, empty circles).
However, when the signal is affected by a gasous environment with either amplitude, the use of the GR+TGR template leads to significant deviations from GR at all PN orders (left panel, blue circles). These false deviations from GR can be avoided by including the environmental effect in the recovery template (right panel), in which case all fractional biases become non-significant.\footnote{Similar trends are seen in the intrinsic-merger parameters, see Fig.~\ref{fig:biasint}.} Another point to note is that all deviations are almost the same in the left panel when not modeling a given effective gas amplitude because  intrinsic-merger parameters are adjusting enough to produce similar $\delta\hat\psi_k$.

\begin{figure}
    \centering
    \includegraphics[width=0.5\textwidth]{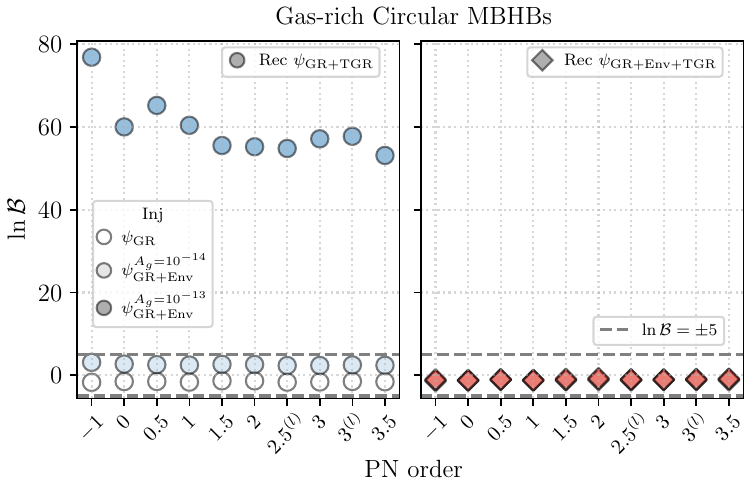}
    \caption{Bayes factors in favour of TGR templates over GR ones, for the analysis of Fig.~\ref{fig:MCMCEnvBias}. We show $\ln\mathcal{B}=\pm5$ (dashed gray lines) for reference. Only the signal with a strong effective gas amplitude (dark filling) induces decisive Bayes factors in favor of the TGR model (left panel). When including the environmental effect in the template (right panel), all $\ln\mathcal{B}$ become inconclusive at approximately $-1.2$, as can be seen with overlapping markers.
    }
    \label{fig:MCMCEnvlnB}
\end{figure}

Similar conclusions can be drawn from the Bayes factors, shown in Fig.~\ref{fig:MCMCEnvlnB}. When we inject the vacuum circular signal, the Bayes factors are inconclusive but favor the GR template (left panel, empty circles). This is to be expected, as both GR and GR+TGR are appropriate templates, with the extra parameter in the latter leading to negative Bayes factors. 
When the injection has an environmental effect with amplitude $A_g=10^{-14}$, despite the significant deviations from GR in terms of fractional biases displayed in Fig.~\ref{fig:MCMCEnvBias}, the Bayes factors only inconclusively favor GR+TGR over GR (left panel, light blue circles). 
This is because our threshold for the Bayes factors is more stringent than a two-sigma bias. 
On the other hand, Bayes factors are strongly in favor of a deviation from GR when the environmental amplitude is larger, $A_g=10^{-13}$. When including the effect of gas in the model (Fig.~\ref{fig:MCMCEnvlnB}, right panel), we  obtain (low-significance) Bayes factors in favor of the GR+Env template over the corresponding TGR template, as they can both correctly fit the injected signal but the latter has an extra free parameter, $\delta\psi_k$.

\subsubsection{Beyond fiducial}\label{S:EnvComp}

In this section, we either consider a different intrinsic-merger parameter among $\{M_z,q,\chi_{1,2},t_c\}$ with respect to our fiducial system, or fix intrinsic-merger parameters altogether. With the former we explore a wider parameter space, while with the latter we explore the effect of having independent information on $\{M_z,q,\chi_{1,2},t_c\}$ from either a merger-ringdown analysis of the same signal or EM counterparts. Different BH spins have a negligible effect on our results, as they affect the signal in the late inspiral and thus do not interact with early-inspiral corrections.

\begin{figure}
    \centering
    \includegraphics[width=0.5\textwidth]{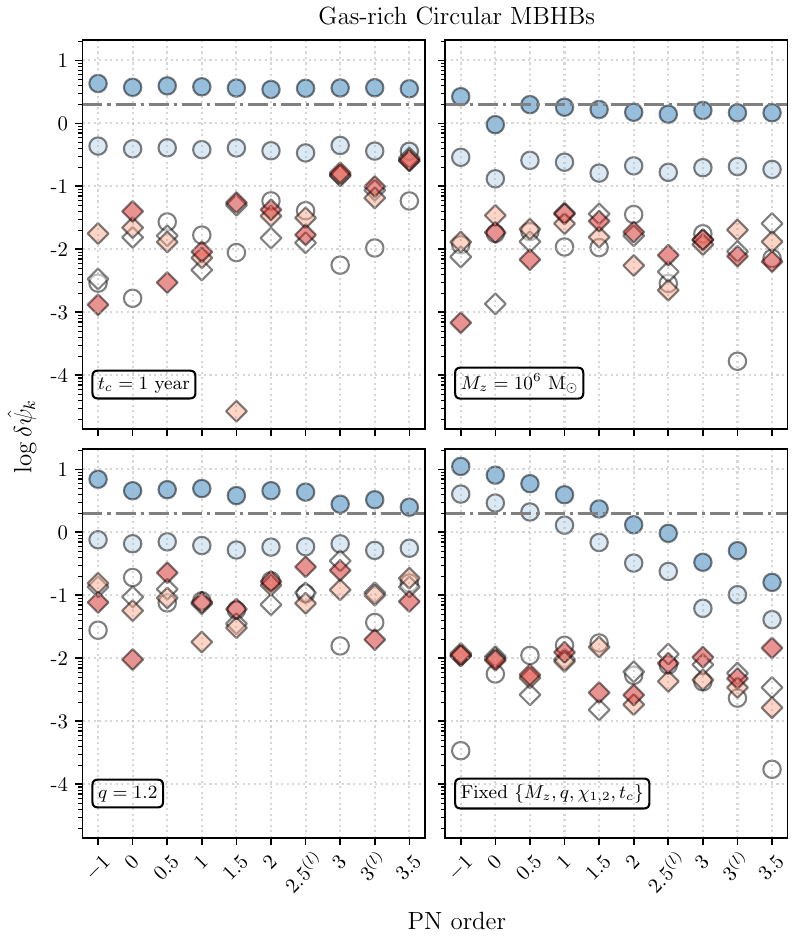}
    \caption{
    Same analysis as in Fig.~\ref{fig:MCMCEnvBias}, but varying one aspect at a time. We reduce $t_c$ to 1 year in the top-left panel, increase $M_z$ to $10^6~\MSun$ in the top-right panel, reduce $q$ to $1.2$ in the bottom-left panel, and assume $\{M_z,q,\chi_{\rm 1,2},t_c\}$ are independently measured in the bottom-right panel. Varying $\{M_z,q,t_c\}$ compared to our fiducial binary leads to less significant deviations from GR, with the smaller environmental amplitude $A_g=10^{-14}$ inducing $\delta\hat\psi_k<2$ at all PN orders. 
    Fixing $\{M_z,q,\chi_{1,2},t_c\}$ leads to significant biases only at lower PN orders $k\lesssim2$, as expected from the early-inspiral nature of the gas effect.
    For the first three systems, \{SNR, no. of GW cycles\} are $\sim\{377,8100\}$, $\sim\{1584,4600\}$, and $\sim\{595,13500\}$, respectively.
    } 
    \label{fig:MCMCEnvComp}
\end{figure}

We show our results in terms of the fractional biases in Fig.~\ref{fig:MCMCEnvComp}. Differences from our fiducial system for a different value of $\{M_z,q,t_c\}$ can be mainly explained by the reduction of detectable GW cycles, which in turn reduces the importance of the early-inspiral environmental effect. Fractional biases are approximately scaled down by the number of expected GW cycles with respect to the fiducial binary. 
The signal SNR plays a minor role here, because the early inspiral typically accounts for $\sim1$-$2$ per cent of the event SNR, which remains large in all cases. These are also the reasons why the smaller environmental amplitude $A_g=10^{-14}$ is not measurable for the variations of $\{M_z,q,t_c\}$ considered here, see \citet{Garg2024b}.

Fixing all intrinsic-merger parameters restricts significant deviations from GR to the lower PN orders, as now $\{M_z,q,\chi_{1,2},t_c\}$ are not able to adjust to give similar $\delta\hat\psi_k$, in contrast to Fig.~\ref{fig:MCMCEnvBias}. See Fig.~\ref{fig:lnBComp} in the Appendix for the corresponding Bayes factors.

\subsection{Vacuum eccentric systems}
We repeat the analysis performed in the last section for an eccentric signal, rather than one affected by gas. We consider two different initial eccentricities $e_0\in\{10^{-2.5},10^{-2.25}\}$. We inject either vacuum circular or vacuum eccentric signals, and recover them with either a GR template with TGR parameters ($\ttgr$) or an eccentric vacuum TGR template ($\tecctgr$). Both values of $e_0$ are measurable by LISA for our fiducial MBHB \citep{Garg2024b}. Finally, we explore the wider parameter space, similarly to Section~\ref{S:EnvComp}.

\subsubsection{Fiducial MBHB}

\begin{figure}
    \centering
    \includegraphics[width=0.5\textwidth]{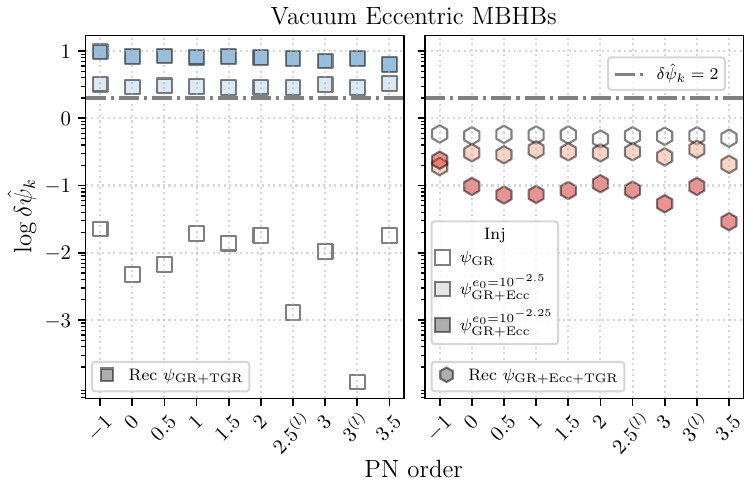}
    \caption{
    Fractional bias induced on TGR parameters by eccentricity. 
    We consider three injections: vacuum circular GR ($\tgr$; no fill), GR with small eccentricity ($e_0=10^{-2.5}$; light fill), and moderate eccentricity ($e_0=10^{-2.25}$; dark fill).  
    The biases indicate significant deviations from GR at all PN orders for either eccentricity when the template neglects the effect ($\ttgr$, circles, left panel). Modelling eccentricity in the TGR template ($\tecctgr$, diamond, right panel) mitigates this systematic bias. 
    See also Fig.~\ref{fig:MCMCEnvBias}.}
    \label{fig:MCMCEccBias}
\end{figure}

In Fig.~\ref{fig:MCMCEccBias}, we show the inferred fractional bias on the TGR parameters. We find that, when injecting a vacuum signal, all $\delta\hat\psi_k<2$ (left panel, empty squares), as expected. Instead, neglecting eccentricity leads to false violations of GR at all PN orders, for either value of the initial eccentricity (left panel, blue squares). However, once we include eccentricity in the TGR waveform, the fractional biases $\delta\hat\psi_k$ become non-significant, implying consistency with GR (right panel). Figure~\ref{fig:biasint} in the Appendix shows the corresponding biases on the intrinsic-merger parameters, which show a similar pattern.

\begin{figure}
    \centering
    \includegraphics[width=0.5\textwidth]{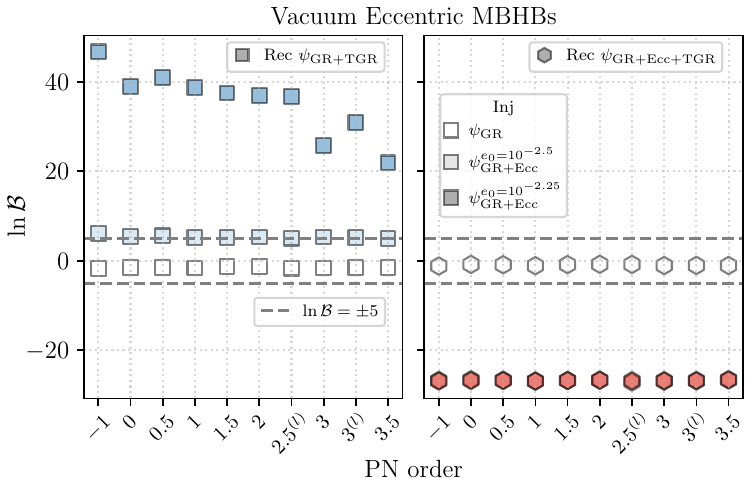}
    \caption{Same as Fig.~\ref{fig:MCMCEnvlnB} but for the eccentric systems considered in Fig.~\ref{fig:MCMCEccBias}. GR+TGR has almost conclusive evidence in its favor over GR for the small eccentricity, and decisive Bayes factors for larger $e_0$ (left panel). Considering eccentricity during Bayesian recovery makes GR+Ecc preferable over GR+Ecc+TGR (right panel), as expected, with overlapping markets for non-zero eccentric cases.
    }
    \label{fig:MCMCEcclnB}
\end{figure}

Similar trends are observed when computing the Bayes factors, Fig.~\ref{fig:MCMCEcclnB}. In the case of a circular GR injection, $\ln\mathcal{B}$ is negative and inconclusive, as expected. For eccentricity $e_0=10^{-2.5}$, we find $\ln\mathcal{B}\sim5$ at all PN orders, while Bayes factor become clearly decisive in favor of a deviation from GR for $e_0=10^{-2.25}$. Including eccentricity in the recovery template leads to Bayes factors slightly favoring the GR hypothesis for $e_0=0$, and strongly favoring the eccentric vacuum template for either eccentricities considered.

\subsubsection{Beyond fiducial}

\begin{figure}
    \centering
    \includegraphics[width=0.5\textwidth]{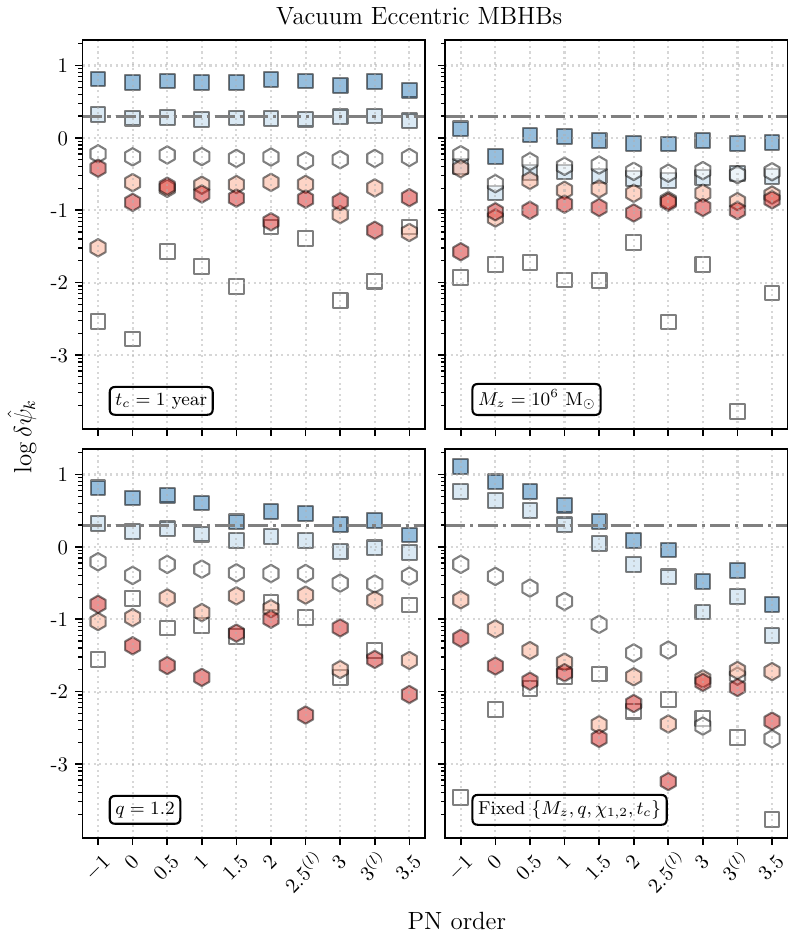}
    \caption{Dependence of the TGR systematics induced by eccentricity (Fig.~\ref{fig:MCMCEccBias}) on the parameters of the MBHB and the analysis. Changing $\{M_z,q,t_c\}$ compared to our fiducial binary reduces the number of detected GW cycles, which in turn reduces the systematic biases. 
    Fixing $\{M_z,q,\chi_{1,2},t_c\}$ in the analysis allows for $\delta\psi_k>2$ only at lower PN orders, those most affected by eccentricity.}
    \label{fig:MCMCEccComp}
\end{figure}

In Fig.~\ref{fig:MCMCEccComp}, we redo the analyses shown in Fig.~\ref{fig:MCMCEnvComp} by changing one aspect of the analysis from our fiducial case at a time. Similar to the case of the environmental effect, differences from our fiducial MBHB are primarily due to the reduction in the number of GW cycles. Again, we observe that violations of GR become less significant for a given eccentricity when changing any of the intrinsic-merger parameters. This is to be expected, as the same eccentricities become unmeasurable in these systems, see \cite{Garg2024c}. We show the corresponding Bayes factors in Fig.~\ref{fig:lnBComp} in the Appendix.

\section{Discussion}\label{S:discussion}
Most current literature, including the LISA Definition Study Report \citep{Colpi2024}, which reflects the primary science goals of the LISA science community, do not consider either orbital eccentricity or gas effects when modelling near-equal mass MBHBs. This is partly due to a lack of fast, efficient, and accurate eccentric waveforms that are appropriate for most eccentricities, and uncertainties in simulating the interaction between a binary and an accretion disc. However, there is also not a general consensus to consider beyond vacuum circular effects due to the expectation that MBHBs will fully circularize due to GWs. There is even less agreement over relevance of environmental effects, due to possibilities of CBDs decoupling much before the binary inspiral in the LISA band or not perturbing the MBHB evolution as strongly as predicted by sub-parsec scale simulations. The latter is hard to verify numerically, because getting a sensible gas torque measurement is extremely hard in the near-merger phase, due to the disparate scales involved and the need for a fully general-relativistic magnetohydrodyanmical (GRMHD) treatment. Therefore, studies in the near-merger phase using GRMHD prescription have focused on EM counterparts \citep{Avara2024,Ennoggi2023,Gutierrez2022}.

In this work, we have systematically illustrated that not modeling even moderate gas-induced perturbations ($A_g=10^{-14}$ or $\fedd\gtrsim0.1$) or a small initial eccentricity ($e_0\gtrsim10^{-2.5}$) induce false, statistically significant, deviations from GR in several PN orders for a fiducial MBHB in LISA. Moreover, unmodelled gas and eccentricity can also bias intrinsic-merger parameters $\{M_z,q,\chi_{\rm 1,2},t_c\}$, which would lead to inconsistencies in inspiral-merger-ringdown tests, as both effects are negligible at and after merger. While Bayes factors are inconclusive in the case of the weakest but measurable non-vacuum effect, i.e., $\fedd=0.1$ in Fig.~\ref{fig:MCMCEnvlnB}, they are strongly in favour of a deviation from GR when the system is eccentric or the gas amplitude larger. 
We find that including these perturbations during Bayesian recovery can mitigate these systematics, while also not significantly affecting the inference of a vacuum circular signal. 

Perhaps unsurprisingly, the parameter space where unmodeled gas effects or eccentricity lead to apparent violations of GR is the same where these effects are measurable (relative error $<50$ per cent) by LISA (see Figs 3 and 4 of \citealt{Garg2024b}). Moreover, the values of $e_0$ or $A_g$ for which GR deviations are significant but Bayes factors are inconclusive are the ones for which the relative measurement error is larger than $30$ per cent \citep{Garg2024b}. 

Concurrent credible deviations from GR and unmodelled gas or eccentricity may not be as problematic as they seem at first glance. This is due to the expectation that beyond GR theories typically manifest themselves at positive PN orders and/or at the merger-ringdown part of the signal \citep{Arun2022}, in contrast to early inspiral effects at negative PN orders considered in this work. Furthermore, any real GR violation should be observed in all MBHB GW events to resolve this confusion further, since the nature of gas-eccentricity systematics is event-specific. Observations of associated EM counterparts can also help to solve this problem.

Our results have consequences for several scientific communities interested in utilizing LISA data \citep{Speri:2022kaq}. Since the fundamental physics community aims to test GR to high accuracy with LISA \citep{Arun2022}, removing gas and eccentricity systematics will be crucial. Biased MBHB parameters are problematic for astrophysicists and cosmologists as well, since they can lead to a bias in the recovered MBH mass function of the Universe. However, probably less than $\sim100\MSun$ bias which is still a better precision then EM observations.
Furthermore, eccentricity and gas imprints are signatures of the binary's formation channel~\citep{Gualandris2022, Franchini:2024pgl}. Lastly, for the data analysis community, the above systematics could impact the ongoing global fit challenge. Although in this work we have not quantified how gas- and eccentricity-induced biases could affect the global fit \citep{Littenberg2023,Katz2024,Strub2024} or population studies with LISA data -- which are both under development -- both are likely to be affected. Because MBHBs will be the loudest LISA source, subtracting them from data with a vacuum circular template could leave residuals, in particular at low frequencies, which could possibly bias the inference of Galactic binaries. Overall, we hope this work motivates the wider LISA community to better understand, and model eccentricity and gas effects when constructing waveform templates for real data in the 2030s.

\subsection{Caveats}

This work has several limitations. We consider aligned-spin corrections, as gas helps in spin alignment, but this will depend on how long gas survives around the MBHB before dissipating or decoupling. We only modeled secular CBD effects, but disks can also induce sub-orbital torque fluctuations that can perturb the binary evolution and which could also be detectable \citep{Zwick2022, Zwick2024}. In addition, CBDs can cause eccentric binary precession, potentially detectable by LISA \citep{Tiede2024b}. We also do not consider the GW-gas coupling that manifests at $-1$PN order in the GW phase recently reported by \citet{Garg2024e}. Since it makes gas effects an order-of-magnitude stronger, we expect that including this additional term should strengthen our conclusion about the need to consider gas perturbations for LISA MBHBs. Other (exotic) environmental effects can also be important for near-equal mass MBHBs, such as those induced by dark matter distributions~\citep{Baumann2019, Alonso-Alvarez2024, Duque:2023seg, Brito:2023pyl, Tomaselli:2024bdd, Kavanagh2024,Bromley2024}. However, these effects are more uncertain, as GWs detected with future detectors are themselves the best way to constrain their existence.

We also neglect higher modes excited by the unequal mass ratio of our fiducial system. The reason behind this approximation is the lack of analytical eccentric and environmental prescriptions for non-dominant modes. We believe that this choice does not impact the main conclusion of our work: having higher modes in the template can help constrain vanilla binary parameters \citep{Pitte2023}, which in turn can enhance false deviations from GR due to mis-modelling. For eccentricity, there are better waveform models \citep{Klein2021,Ramos-Buades2023,Ramos-Buades:2023yhy}, which are accurate beyond $e_0=0.2$. However, it is unclear how much worse \textsc{TaylorF2Ecc} performs against these better templates for eccentricities below $0.01$. Moreover, in this study, we fix the power-law of the gas-induced dephasing to $n_g=4$, which may not be the case in reality. This should motivate more accurate numerical simulations of binary-disc interactions. Lastly, we do not consider both gas effects and eccentricity together for two reasons. First, they can interact with each other and make the overall non-GR effect stronger or weaker. Second, it can reduce the effectiveness of the crucial point we made in this study. One would expect false deviations for the same gas-eccentricity effects, which are measurable by LISA for a given MBHB \citep{Garg2024b}.

\section{Conclusion}\label{S:conclusion}
In this paper, we studied near-equal mass, aligned-spin MBHBs with total redshifted mass $M_z\sim[10^4,10^7]~\MSun$ in their final year(s) of inspiral in the LISA band before the merger. These binaries are primarily evolving due to the GW radiation expected from a circular binary in vacuum. We considered only the inspiral part of the signal (until the ISCO) with up to $3.5$PN order accuracy in the GW phase. We then introduced small perturbations to the signal under the SPA, either due to the presence of a CBD or a small initial orbital eccentricity $e_0$. The former manifests itself as a -$4$PN order dephasing with an effective gas-amplitude $A_g$ that depends only on the disc properties, including the accretion rate $\fedd$. For eccentricity, we considered corrections up to 3PN order and $\mathcal{O}(e^2)$. 

We performed a parameterized test of GR to assess whether ignoring a non-vacuum or non-circular effect present in the signal can mimic a violation of GR. To achieve this, we added TGR parameters to our templates, $\delta\psi_k$ at the $k/2$-th PN order, including at $-1$PN and $0.5$PN orders. 
To reliably mimic an analysis of LISA data, we included a model of the complete LISA response, inferred all the intrinsic parameters together, and explored the relevant parameter space. For non-GR effects, we considered $A_g\in\{10^{-14},10^{-13}\}$ and $e_0\in\{10^{-2.5},10^{-2.25}\}$. To be a statistically significant (apparent) deviation from GR, we required either more than two-sigma confidence in a non-zero value of $\delta\psi_k$, i.e. fractional bias $\delta\hat\psi_k>2$, or the natural logarithm of the Bayes factor ($\ln\mathcal{B}$) in favor of the TGR over the GR hypothesis to be more than $5$. 

Here we list the key takeaways from this work:
\begin{itemize}
    \item For our fiducial MBHB ($M_z=10^5~\MSun,~q=8,~\chi_{1,2}=0.9$, and $t_c=4$ years, see the remaining parameters in Table~\ref{table:parameters}),
    unmodelled eccentricity or gas lead to a false violation of GR at all PN orders in terms of biases $\delta\hat\psi_k>2$ (see Figs.~\ref{fig:MCMCEnvBias} and \ref{fig:MCMCEccBias}).
    \item Unmodelled eccentricity or gas also induce biases in the intrinsic-merger parameters $\{M_z,q,\chi_{\rm 1,2},t_c\}$ (see Fig.~\ref{fig:biasint}).
    \item Bayes factors for deviations from GR are decisive 
    for both values of eccentricity, and but only for the stronger gas amplitude, as $\ln\mathcal{B}>5$ is a more stringent requirement than $\delta\hat\psi_k>2$ (see Figs~\ref{fig:MCMCEnvlnB} and \ref{fig:MCMCEcclnB}). 
    \item Including the ignored beyond vacuum-circular effects mitigates these systematics, while also not biasing TGR parameters of a signal that does not have either perturbation (see Figs~\ref{fig:MCMCEnvBias}, \ref{fig:MCMCEnvlnB}, \ref{fig:MCMCEccBias}, and \ref{fig:MCMCEcclnB}).
    \item Beyond our fiducial binary, changing to either $M_z=10^6~\MSun$, $q=1.2$, or $t_c=1$ years  leads to less significant deviations from GR, due to a reduction in the number of GW cycles (see Figs~\ref{fig:MCMCEnvComp} and \ref{fig:MCMCEccComp}). 
    \item Finally, if we fix the intrinsic-merger parameters during Bayesian inference under the assumption that they are known from the merger-ringdown of the same signal or EM counterparts, false deviations from GR are limited to lower PN orders
    (see Figs~\ref{fig:MCMCEnvComp} and \ref{fig:MCMCEccComp}).
\end{itemize}

\section*{Data availability statement}

The data underlying this article will be shared on reasonable request to the authors.

\section*{Acknowledgements}
M.G.~acknowledges support from the Swiss National Science Foundation (SNSF) under the grant 200020\_192092. L.S. acknowledges support from the UKRI guarantee funding (project no. EP/Y023706/1). We acknowledge John G.~Baker and Sylvain Marsat for providing us access to their \textsc{lisabeta} software. We also thank Cecilia Chirenti for useful discussion. The authors also acknowledge use of NumPy  \citep{harris2020array} and inspiration drawn from the \textsc{GWFAST} package \citep{Iacovelli2022} regarding the python implementation of \textsc{TaylorF2Ecc}.

\scalefont{0.94}
\setlength{\bibhang}{1.6em}
\setlength\labelwidth{0.0em}
\bibliographystyle{mnras}
\bibliography{TGR}
\normalsize

\appendix
\section{Bayesian setup}\label{App:Bayesian}

We use zero-noise realization, meaning that while we do weigh our waveforms with LISA's power spectral density, we do not add any additional noise to the injected signal. The expectation is that a realization of noise should produce similarly shaped posteriors with a slight bias in their median.

When using Fisher initialization, we draw random samples from a multivariate Gaussian of the parameters of interest with mean given by the injected values and covariance computed with the Fisher formalism \citep{Vallisneri2008}. This speeds up the likelihood computations based on the reasonable assumption that a near-equal mass MBHB with low eccentricity should have almost all its power in the primary mode.

In Table~\ref{table:priors}, we show our priors on system parameters. For the TGR parameters, we take their best constrained values from LVK observations as priors \citep{TGR_GWTC3}.
\begin{table}
\centering
    \begin{tabular}{|C|C|}
        \hline
        {\rm Parameter}&{\rm Uniform \ priors}\\
        \hline
        \hline
        M_z[\MSun]&[10^4,10^8]\\
        \hline
        q&[1,10]\\
        \hline
        \chi_{1,2}&[-1,1]\\
        \hline
        t_c[{\rm yr}]&t_{c,\rm inj}[{\rm yr}]+[-0.5,0.5]\\
        \hline
        e_0&[10^{-6},0.2]\\
        \hline
        A_g[10^{-15}]&[-10^4,10^4]\\
        \hline
        \delta\phi_{-2}&[-10^{-4},10^{-4}]\\
        \hline
        \delta\phi_0~{\rm or}~\delta\phi_1&[-10^{-1},10^{-1}]\\
        \hline
        \delta\phi_k~\forall~k>1&[-1,1]\\
        \hline
    \end{tabular}
\caption{Uniform Priors on all parameters of interest.}
\label{table:priors}
\end{table}
\section{Free vs Fixed phase at coalescence}\label{App:phic}

To show that there is a partial degeneracy between higher order TGR coefficients ($\delta\phi_k$) and the phase at coalescence ($\phi_c$), we employ the Fisher formalism \citep{Vallisneri2008}. We compute the Fisher matrix for our waveform model $\tilde{h}(f)$, by only considering $\phi_c$ and $\delta\phi_k$ as free parameters. The vacuum circular waveform containing a single TGR parameter with index $k$ can be written as
\begin{align}
    \tilde{h}(f)&=Ae^{{i\psi_{\rm TGR,k}}},\\ 
    {\rm where}~&\psi_{\rm TGR,k}=2\pi ft_c-\phi_c+\nonumber\\
    &\frac{3}{128\eta v^{5}}\left[\sum_{n}\left(\psi_{n}+\psi_{n}^{(l)}\log v\right)v^{n}+\psi_{k}\delta\psi_kv^k\right].\nonumber
\end{align}
Then we can compute the partial derivative of $\tilde{h}$ as a function of either $\phi_c$ or $\delta\phi_k$:
\begin{align}
    \frac{\partial \tilde h}{\partial \phi_c}=-i\tilde h,~{\rm and}~\frac{\partial \tilde h}{\partial \delta\psi_k}=i\frac{3}{128\eta v^{5}}\psi_{k}v^k\tilde h.
\end{align}
Therefore, the elements of the Fisher matrix are
\begin{align}\label{eq:Fisher}
    \left(\frac{\partial \tilde h}{\partial \phi_c}\bigg|\frac{\partial \tilde h}{\partial \phi_c}\right)&=-(\tilde h| \tilde h),\\
    \left(\frac{\partial \tilde h}{\partial \phi_c}\bigg|\frac{\partial \tilde h}{\partial \delta\psi_k}\right)&=\frac{3}{128\eta}\psi_k(\tilde h| v^{k-5} \tilde h)\nonumber,\\
    \left(\frac{\partial \tilde h}{\partial \delta\psi_k}\bigg|\frac{\partial \tilde h}{\partial \delta\psi_k}\right)&=-\left(\frac{3}{128\eta}\psi_k\right)^2( v^{k-5} \tilde h| v^{k-5} \tilde h)\nonumber,
\end{align}
where $(\cdot|\cdot)$ is the standard inner product and $(\tilde h| \tilde h)$ is simply the SNR$^2$. 

Now its apparent from Eq.~\eqref{eq:Fisher} that for $k=5$, the Fisher matrix is singular (its determinant is zero). This implies a full degeneracy between the phase at coalescence and the $k^{\rm th}$ TGR parameter. Even for $k\sim5$, there is a partial degeneracy between $\phi_c$ and $\delta\psi_k$, which is exacerbated by the limited number of GW cycles at those higher PN orders. As a result, higher TGR parameters behave as quasi-monochromatic. 
\section{Biases on intrinsic merger parameters factors}\label{App:Bias}
\begin{figure}
    \centering
    \includegraphics[width=0.45\textwidth]{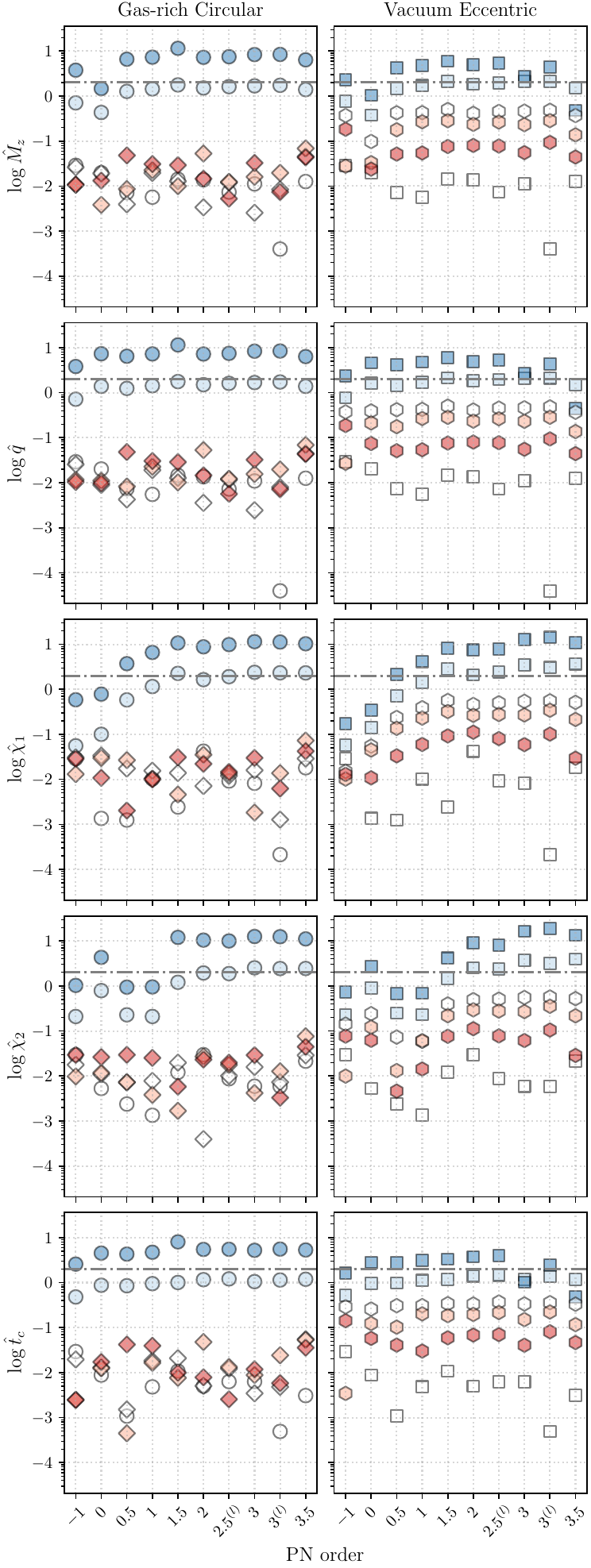}
    \caption{Fractional biases on intrinsic-merger parameters $\{M_z,q,\chi_1,\chi_2,t_c\}$ for our fiducial MBHB from top to bottom panels, respectively. The left panels shows the biases induced in a TGR waveform model when the injection is affected by gas, as in Fig.~\ref{fig:MCMCEnvBias}. The right panels show the orbital eccentricity case, as in Fig.~\ref{fig:MCMCEccBias}.}
    \label{fig:biasint}
\end{figure}

In Fig.~\ref{fig:biasint}, we show the fractional biases on the intrinsic-merger parameters corresponding to the study of violations of GR for our fiducial MBHB in Figs~\ref{fig:MCMCEnvBias} and \ref{fig:MCMCEccBias}. Similarly to the case of TGR parameters, including the appropriate non-vacuum or non-circular effect during Bayesian inference makes fractional biases on $\{M_z,q,\chi_1,\chi_2,t_c\}$ parameters non-significant.
\section{Bayes factors}\label{App:Bayes}
We show the Bayes factors for the different scenarios considered in Fig.~\ref{fig:MCMCEnvComp} and Fig.~\ref{fig:MCMCEccComp}. The Bayes factors follow the same pattern as the fractional bias: $|\ln\mathcal{B}|\lesssim5$ indicate non-conclusive evidence in favour of a deviation from GR once the unmodeled environmental or eccentric correction is included in the GW waveform. Since, we can not recover with GR model when fixing intrinsic-merger parameters, we naturally do not have circles or squares in the bottom two panels. 
\begin{figure}
    \centering
    \includegraphics[width=0.5\textwidth]{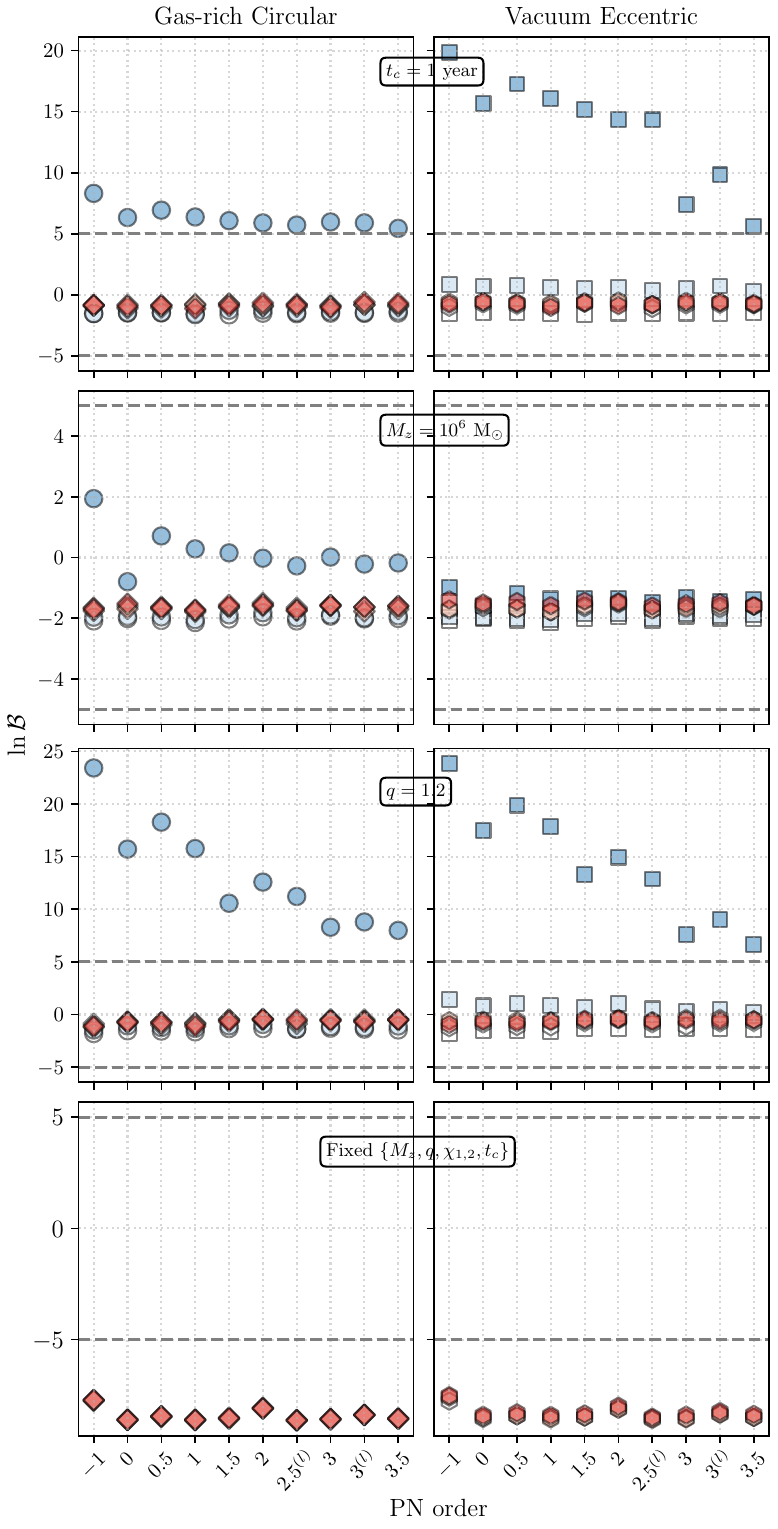}
    \caption{Bayes factors corresponding to Fig.~\ref{fig:MCMCEnvComp} systems on the left and  Fig.~\ref{fig:MCMCEccComp} systems on the right.}
    \label{fig:lnBComp}
\end{figure}

\bsp 
\label{lastpage}
\end{document}